\documentclass[proceedings]{JHEP3}
\usepackage{amssymb}
\usepackage{amsmath}
\usepackage{epsfig,multicol}

\newbox\mybox

\newcommand\fverb{\setbox\mybox=\hbox\bgroup\verb}
\newcommand\fverbdo{\egroup\medskip\noindent\fbox{\unhbox\mybox}\ }
\newcommand\fverbit{\egroup\item[\fbox{\unhbox\mybox}]}
\conference{$G_2$-Calogero-Moser Lax operators from reduction}
\abstract{We construct a Lax operator for the $G_2$-Calogero-Moser model
by means of a double reduction procedure. In the first reduction step we reduce
the $A_6$-model to a $B_3$-model with the help of an embedding of the $B_3$-root 
system into the $A_6$-root system together with the specification of certain 
coupling constants. The $G_2$-Lax operator is obtained thereafter by means 
of an additional reduction by exploiting the embedding of the $G_2$-system 
into the $B_3$-system. The degree of algebraically independent and non-vanishing 
charges is found to be equal to the degrees of the corresponding Lie algebra.}

\title{$G_2$-Calogero-Moser Lax operators from reduction}
\author{Andreas Fring$^\bullet$ and Nenad Manojlovi\'c$^\circ$ \\
$^\bullet$  Centre for Mathematical Science, City University, \\
$\;$ Northampton Square, London EC1V 0HB, UK\\
$\;$ E-mail: \email{A.Fring@city.ac.uk}\\
$^\circ$ Departamento de Matem\'{a}tica, F.C.T. Universidade do Algarve,\\
$\;$ Campus de Gambelas, 8005-139 Faro, Portugal\\
$\;$ E-mail: \email{nmanoj@ualg.pt}}

\input{tcilatex}

\begin{document}

\section{Introduction}

The Calogero-Moser models \cite{Cal1,Cal2,Cal3,Suth3,Suth4,Mo,OP2,Per}
constitute a large class of well studied interacting many particle systems.
The models are very universal in the sense that they can be cast into a form
in which the potential term includes a sum over all roots $\alpha $ of some
root system $\Delta $ and the functional dependence of the potential is $%
V(x)\sim 1/\func{sn}^{2}(x)$, with $\func{sn}$ being an elliptic function
together with its various limits $1/\sinh ^{2}(x)$, $1/\sin ^{2}(x)$ and $%
1/x^{2}$. Often it is useful to treat the latter cases independently for
their own sake. Due to their universal nature the models find a wide range
of applications in physics, as for instance to characterize anyons on the
lowest Landau level \cite{Brink}, to describe certain properties of quantum
Hall droplets \cite{IR} and in various ways in conformal \cite%
{CFT1,CFT2,CFT3,CFT4} and boundary \cite{Cardy} conformal field theories.

The Hamiltonian for an $n$-particle Calogero-Moser system reads 
\begin{equation}
\mathcal{H}=\frac{p^{2}}{2}-\frac{1}{2}\sum\limits_{\alpha \in \Delta
}g_{\alpha }^{2}V(\alpha \cdot q)\qquad g_{\alpha }\in i\mathbb{R},\quad
q,p\in \mathbb{R}^{n},  \label{H}
\end{equation}
with $n$ being the dimensionality of the space in which the roots $\alpha $
are realized. At this point we impose only the restrictions $g_{\alpha
}=g_{-\alpha }$ on the coupling constant, even though later on we equate
more of them for reasons to be explained. One of the most prominent feature
of these models is their integrability, meaning here the existence of a
sufficient number of conserved quantities (integrals of motion) $I_{k}$ in
involution. A standard technique to construct these charges, the so-called
isospectral deformation method, goes back almost forty years \cite{Lax}. It
consists of formulating Lax pair operators $L$ and $M$ as functions of the
dynamical variables $q_{i}$ and $p_{i}$ for $1\leq i\leq n$, which satisfy
the Lax equation $\dot{L}=\left[ L,M\right] $, upon the validity of the
classical equation of motion resulting from (\ref{H}). The Lax operator is
then the starting point for the construction of conserved charges of the
form $I_{k}=\func{tr}(L^{k})/k$, with $I_{2}\sim \mathcal{H}$, of classical
r-matrices \cite{STS,Bab,Avan,Avan2,Skly,braden}, spectral curves \cite%
{Krich,BB,Don,Hok} and various other important quantities. For root systems $%
\Delta $, which can be associated with a Lie algebra \textbf{g}, i.e.
crystallographic ones\footnote{%
For non-crystallographic root systems one may exploit the fact that they are
embedded into crystallographic ones and use a reduction procedure to obtain
a meaningful Lie algebraic operator \cite{FK}.}, a natural Ansatz is to
expand $L$ and $M$ in terms of the elements $H,E_{\alpha }$ of \textbf{g} 
\begin{equation}
L=p\cdot H+\sum\limits_{\alpha \in \Delta }g_{\alpha }f(\alpha \cdot
q)E_{\alpha }\quad \text{and\quad }M=m\cdot H+\sum\limits_{\alpha \in \Delta
}g_{\alpha }h(\alpha \cdot q)E_{\alpha }.  \label{LM}
\end{equation}
Alternatively, one may also expand $L$ and $M$ in terms of other
non-commuting objects, such as Coxeter transformations, and perform a
similar analysis \cite{Sas6,Sas5}. Substitution of these operators into the
Lax equation yields various constraining equations, which for a given
potential determine the functions $f(x)$ and $h(x)$ in $L$ and $M$ as
defined in equation (\ref{LM}). We choose here as convention the Cartan-Weyl
basis $\func{tr}(H_{i}H_{j})=\delta _{ij}$, $\func{tr}(E_{\alpha }E_{-\alpha
})=1$, which is consistent with the well-known commutation relations (e.g. 
\cite{Hum}) 
\begin{equation}
\left[ H_{i},H_{j}\right] =0,~~~\left[ H_{i},E_{\alpha }\right] =\alpha
^{i}E_{\alpha },~~~\left[ E_{\alpha },E_{-\alpha }\right] =\alpha \cdot H,~~~%
\left[ E_{\alpha },E_{\beta }\right] =\varepsilon _{\alpha ,\beta }E_{\alpha
+\beta }.  \label{comm}
\end{equation}
Then by direct substitution it follows that the Lax equation holds once the
functions $f_{\alpha }(x)=g_{\alpha }f(x)$ and $h_{\alpha }(x)=g_{\alpha
}h(x)$ satisfy 
\begin{equation}
g(x)=f^{\prime }(x),\quad \dot{p}=\sum\limits_{\alpha \in \Delta }\alpha
f_{\alpha }(\alpha \cdot q)h_{-\alpha }(-\alpha \cdot q),\quad \gamma \cdot
m=\sum\limits_{\substack{ \alpha ,\beta \in \Delta  \\ \alpha +\beta =\gamma 
}}\varepsilon _{\alpha ,\beta }\frac{f_{\alpha }(\alpha \cdot q)h_{\beta
}(\beta \cdot q)}{f_{\gamma }(\gamma \cdot q)}.  \label{C1}
\end{equation}
Assuming further that $I_{2}=\mathcal{H}$ and the classical equation of
motions resulting from (\ref{H}), one obtains three additional equations 
\begin{equation}
f(x)f(-x)=-V(x),\quad \dot{p}=-\frac{1}{2}\sum\limits_{\alpha \in \Delta
}\alpha g_{\alpha }^{2}V^{\prime }(\alpha \cdot q)\quad \text{and\quad }%
f(x)=\pm f(-x).  \label{C2}
\end{equation}
For the stated potentials it is straightforward to use the factorizing
condition in (\ref{C2}) to determine $f(x)$, i.e. $1/\func{sn}(x)$, $1/\sinh
(x)$, $1/\sin (x)$ and $1/x,$ and therefore $g(x)$ by taking its derivative.
Thus to establish the integrability of the system (\ref{H}) reduces to the
question of whether the third set of equations in (\ref{C1}) admits a
solution for the vector $m$ and therefore guarantees the existence of the
operator $M$. As these equations are in general highly overdetermined, the
answer to this question depends crucially on the structure of the Lie
algebra. As was observed long time ago \cite{Func}, for given potentials as
in (\ref{H}) the relation 
\begin{equation}
f(x)f^{\prime }(y)-f^{\prime }(x)f(y)=f(x+y)\left[ V(x)-V(y)\right]
\label{f}
\end{equation}
holds, such that the equations may be simplified further. Then the last set
of equations in (\ref{C1}) reduces to 
\begin{equation}
g_{\gamma }(\gamma \cdot m)=\sum\limits_{\substack{ \alpha ,\beta \in \Delta 
\\ \alpha +\beta =\gamma }}\varepsilon _{\alpha ,\beta }g_{\alpha }g_{\beta
}V(\alpha \cdot q)=\sum\limits_{\alpha \in \Delta }\varepsilon _{\alpha
,\gamma }g_{\alpha }g_{\alpha +\gamma }V(\alpha \cdot q).  \label{V}
\end{equation}
Clearly 
\begin{equation}
m=\sum\limits_{\alpha \in \Delta }\sum\limits_{i=1}^{\ell }\varepsilon
_{\alpha ,\gamma _{i}}\frac{g_{\alpha }g_{\alpha +\gamma _{i}}}{g_{\gamma
_{i}}}V(\alpha \cdot q)\lambda _{i},  \label{m}
\end{equation}
with $\lambda _{i}$ being a fundamental weight, is a solution to (\ref{V})
when $\gamma _{i}$ is taken to be a simple root. However, for (\ref{m}) to
be a proper solution one also has to verify whether it solves (\ref{V}) for
the remaining roots. In summary, we can say that if the system (\ref{V}) can
be solved for the vector $m$ for a particular Lie algebra \textbf{g}, then
the system (\ref{H}) is classically integrable. The reverse statement does
not hold.

\section{The $G_{2}$-Lax operator}

It turns out that only when the algebra \textbf{g} in (\ref{LM}) is taken to
be $A_{\ell }$ one obtains directly, meaning that all quantities in (\ref{H}%
) belong to $A_{\ell }$, a solution for the Lax operator with the condition
that the corresponding equation of motion holds. In all other cases one
needs to device alternative methods. The expressions for the $B_{\ell }$, $%
C_{\ell }$ and $D_{\ell }$-algebras were obtained \cite{OP6,OP2,Per} from $%
A_{2\ell }$-theories by specific transformations of the dynamical variables
and a subsequent constraint on certain coupling constants $g_{\alpha }$. For
the remaining algebras different types of techniques have been developed 
\cite{OP6,OP2,Per,Hok,Sas6,Sas5}. Surprisingly for the Lie algebras $%
E_{6,7,8},F_{4}$ and $G_{2}$ no Lax pair was known until fairly recent \cite%
{Hok}.

In particular, the latter, the $G_{2}$-Calogero-Moser model, constitutes a
standard simple example, since it can be viewed as the classical three-body
problem with a two and a three-body interaction term \cite{Wolf}. For a
specific realization of the roots (see below), the potential term in the
rational case simply reads 
\begin{equation}
V(\tilde{q})=\frac{\tilde{g}_{s}^{2}}{2}\dsum\limits_{1\leq i<j\leq 3}\frac{1%
}{(\tilde{q}_{i}-\tilde{q}_{j})^{2}}+\frac{\tilde{g}_{l}^{2}}{2}\dsum\limits 
_{\substack{ 1\leq i<j\leq 3  \\ i,j\neq k}}\frac{1}{(\tilde{q}_{i}+\tilde{q}%
_{j}-2\tilde{q}_{k})^{2}}~,  \label{v2}
\end{equation}
with $\tilde{g}_{s},\tilde{g}_{l}$ being coupling constants. It appears to
be rather surprising that despite the simplicity of this model, apart from
the expressions in \cite{Hok}, a general simple formula for the Lax operator
along the line of the original work of \cite{OP6,OP2,Per} may not be found
in the literature. It will be the purpose of this paper to provide such a
simple expression.

\subsection{Direct computation}

Let us commence by directly analyzing equation (\ref{V}) for $G_{2}$. For
the explicit calculation we require first the roots of $G_{2}$. We recall
the general fact, see e.g. \cite{Hum2,PD1,FO}, that the entire root
system can be generated by $h-1$ successive actions of the Coxeter element $%
\sigma $ on bi-coloured simple roots, i.e. $\gamma =\pm $ $\alpha _{i}$,
with $h$ being the Coxeter number. It turns out to be convenient to
abbreviate the roots accordingly, that is we define $\sigma ^{p}\gamma
_{i}=:\alpha _{i,p}$ for $1\leq i\leq \ell =$rank \textbf{g }and $0\leq
p\leq h-1$.

For $G_{2}$ we have $\ell =2$, $h=6$ and the $2\times 6=12$ roots are
computed to

\begin{center}
$ 
\begin{tabular}{|c||c|c|c|c|c|c|}
\hline
$i\diagdown p$ & 0 & 1 & 2 & 3 & 4 & 5 \\ \hline\hline
1 & \multicolumn{1}{r|}{$-\tilde{\alpha}_{1}$} & \multicolumn{1}{r|}{$-(2%
\tilde{\alpha}_{1}+\tilde{\alpha}_{2})$} & \multicolumn{1}{r|}{$-(\tilde{%
\alpha}_{1}+\tilde{\alpha}_{2})$} & \multicolumn{1}{r|}{$\tilde{\alpha}_{1}$}
& \multicolumn{1}{r|}{$2\tilde{\alpha}_{1}+\tilde{\alpha}_{2}$} & 
\multicolumn{1}{r|}{$\tilde{\alpha}_{1}+\tilde{\alpha}_{2}$} \\ \hline
2 & \multicolumn{1}{r|}{$\tilde{\alpha}_{2}$} & \multicolumn{1}{r|}{$-(3%
\tilde{\alpha}_{1}+\tilde{\alpha}_{2})$} & \multicolumn{1}{r|}{$-(3\tilde{%
\alpha}_{1}+2\tilde{\alpha}_{2})$} & \multicolumn{1}{r|}{$-\tilde{\alpha}%
_{2} $} & \multicolumn{1}{r|}{$3\tilde{\alpha}_{1}+\tilde{\alpha}_{2}$} & 
\multicolumn{1}{r|}{$3\tilde{\alpha}_{1}+2\tilde{\alpha}_{2}$} \\ \hline
\end{tabular}
\medskip $

The roots $\hat{\alpha}_{i,p}$ of the $\hat{\Delta}_{G_{2}}$-root system.
\end{center}

\noindent In addition, we require the structure constants $\varepsilon
_{\alpha ,\beta }$ for the analysis of (\ref{V}). The square of the latter
can be fixed by means of the well-known formula $\varepsilon _{\alpha ,\beta
}^{2}=\alpha ^{2}n(m+1)/2$, where the integers $n,m$ are determined by the
so-called $\alpha $-string through $\beta $, i.e. the largest values for $%
n,m $ such that $\beta +n\alpha $ and $\beta -m\alpha $ are still roots, see
e.g. \cite{Bou}. The overall signs are in general not fixed and are subject
to convention. However, some consistency relations have to hold, resulting
from the anti-symmetry of the commutator, the reality condition and the
Jacobi identity when $\alpha +\beta =\gamma $%
\begin{eqnarray}
\varepsilon _{\alpha ,\beta } &=&-\varepsilon _{\beta ,\alpha }=\varepsilon
_{\beta ,-\gamma }=-\varepsilon _{-\gamma ,\beta }=-\varepsilon _{\alpha
,-\gamma }=\varepsilon _{-\gamma ,\alpha }=  \label{eps1} \\
-\varepsilon _{-\alpha ,-\beta } &=&\varepsilon _{-\beta ,-\alpha
}=-\varepsilon _{-\beta ,\gamma }=\varepsilon _{\gamma ,-\beta }=\varepsilon
_{-\alpha ,\gamma }=-\varepsilon _{\gamma ,-\alpha }.  \label{eps2}
\end{eqnarray}
We choose here the short roots $\tilde{\alpha}_{1,p}$ to have length $\tilde{%
\alpha}^{2}=2$ and the long roots $\tilde{\alpha}_{2,p}$ to have length $%
\tilde{\alpha}^{2}=6$. As complete lists of structure constants are
difficult to find in the literature, we present here a consistent choice for
the $12\times 12=144$ structure constants, with 60 of them non-vanishing

\begin{center}
\begin{tabular}{|r||r|r|r|r|r|r|r|r|r|r|r|r|}
\hline
$\tilde{\alpha}_{i,p}\diagdown \tilde{\alpha}_{j,q}$ & $\tilde{\alpha}_{1,0}$
& $\tilde{\alpha}_{1,1}$ & $\tilde{\alpha}_{1,2}$ & $\tilde{\alpha}_{1,3}$ & 
$\tilde{\alpha}_{1,4}$ & $\tilde{\alpha}_{1,5}$ & $\tilde{\alpha}_{2,0}$ & $%
\tilde{\alpha}_{2,1}$ & $\tilde{\alpha}_{2,2}$ & $\tilde{\alpha}_{2,3}$ & $%
\tilde{\alpha}_{2,4}$ & $\tilde{\alpha}_{2,5}$ \\ \hline\hline
\multicolumn{1}{|c||}{$\tilde{\alpha}_{1,0}$} & $0$ & $\mu $ & $-2$ & $0$ & $%
2$ & $\mu $ & $0$ & $0$ & $0$ & $-\mu $ & $-\mu $ & $0$ \\ \hline
\multicolumn{1}{|c||}{$\tilde{\alpha}_{1,1}$} & $-\mu $ & $0$ & $\mu $ & $2$
& $0$ & $-2$ & $0$ & $0$ & $0$ & $0$ & $\mu $ & $-\mu $ \\ \hline
\multicolumn{1}{|c||}{$\tilde{\alpha}_{1,2}$} & $2$ & $-\mu $ & $0$ & $\mu $
& $-2$ & $0$ & $-\mu $ & $0$ & $0$ & $0$ & $0$ & $\mu $ \\ \hline
\multicolumn{1}{|c||}{$\tilde{\alpha}_{1,3}$} & $0$ & $-2$ & $-\mu $ & $0$ & 
$-\mu $ & $2$ & $\mu $ & $\mu $ & $0$ & $0$ & $0$ & $0$ \\ \hline
\multicolumn{1}{|c||}{$\tilde{\alpha}_{1,4}$} & $-2$ & $0$ & $2$ & $\mu $ & $%
0$ & $-\mu $ & $0$ & $-\mu $ & $\mu $ & $0$ & $0$ & $0$ \\ \hline
\multicolumn{1}{|c||}{$\tilde{\alpha}_{1,5}$} & $-\mu $ & $2$ & $0$ & $-2$ & 
$\mu $ & $0$ & $0$ & $0$ & $-\mu $ & $\mu $ & $0$ & $0$ \\ \hline
\multicolumn{1}{|c||}{$\tilde{\alpha}_{2,0}$} & $0$ & $0$ & $\mu $ & $-\mu $
& $0$ & $0$ & $0$ & $0$ & $-\mu $ & $0$ & $\mu $ & $0$ \\ \hline
\multicolumn{1}{|c||}{$\tilde{\alpha}_{2,1}$} & $0$ & $0$ & $0$ & $-\mu $ & $%
\mu $ & $0$ & $0$ & $0$ & $0$ & $\mu $ & $0$ & $-\mu $ \\ \hline
\multicolumn{1}{|c||}{$\tilde{\alpha}_{2,2}$} & $0$ & $0$ & $0$ & $0$ & $%
-\mu $ & $\mu $ & $\mu $ & $0$ & $0$ & $0$ & $\mu $ & $0$ \\ \hline
\multicolumn{1}{|c||}{$\tilde{\alpha}_{2,3}$} & $\mu $ & $0$ & $0$ & $0$ & $%
0 $ & $-\mu $ & $0$ & $-\mu $ & $0$ & $0$ & $0$ & $\mu $ \\ \hline
\multicolumn{1}{|c||}{$\tilde{\alpha}_{2,4}$} & $\mu $ & $-\mu $ & $0$ & $0$
& $0$ & $0$ & $-\mu $ & $0$ & $\mu $ & $0$ & $0$ & $0$ \\ \hline
\multicolumn{1}{|c||}{$\tilde{\alpha}_{2,5}$} & $0$ & $\mu $ & $-\mu $ & $0$
& $0$ & $0$ & $0$ & $\mu $ & $0$ & $-\mu $ & $0$ & $0$ \\ \hline
\end{tabular}
\smallskip

The $G_{2}$-structure constants $\varepsilon _{i,p,j,q}$ with $\mu =\sqrt{3}$%
.
\end{center}

In order to obtain the previous table we only fixed five signs by convention
and determined the remaining ones by means of the relations (\ref{eps1}) and
(\ref{eps2}).

Having assembled all necessary data, we can present a simple argument which
demonstrates that it is not possible to solve the constraint (\ref{V})
directly. Choosing all coupling constants $g_{\tilde{\alpha}}$ to be either $%
g_{s}$ or $g_{l}$ for $\tilde{\alpha}$ to be a short or long root,
respectively, we may for instance add up the equation (\ref{V}) for the
three choices $\gamma =\tilde{\alpha}_{1,0}$, $\gamma =\tilde{\alpha}_{1,2}$
and $\gamma =\tilde{\alpha}_{1,4}$ and find after cancellation of $g_{s}$ 
\begin{equation*}
(\tilde{\alpha}_{1,0}+\tilde{\alpha}_{1,2}+\tilde{\alpha}_{1,4})\cdot m=0=2%
\sqrt{3}g_{l}\left[ V(\tilde{\alpha}_{1,0}\cdot q)-V(\tilde{\alpha}%
_{2,0}\cdot q)+V(\tilde{\alpha}_{1,4}\cdot q)-V(\tilde{\alpha}_{2,4}\cdot q)%
\right] .
\end{equation*}%
Clearly, the right hand side is not zero in general, and hence we can not
solve the constraint (\ref{V}) directly. One reaches the same conclusion by
taking the expression in (\ref{m}) and trying to verify (\ref{V}) for $%
\gamma $ to be a non-simple root.

However, if we switch off the two particle interaction, i.e. we take $%
g_{s}=0 $ in (\ref{v2}), we may construct a particular solution. Taking for
instance $\tilde{\alpha}_{1}=\varepsilon _{1}-\varepsilon _{2}$, $\tilde{%
\alpha}_{2}=-2\varepsilon _{1}+\varepsilon _{2}+\varepsilon _{2}$ as
concrete realization for the simple roots of $G_{2}$ in $\mathbb{R}^{3}$,
with $\varepsilon _{i}\cdot \varepsilon _{j}=\delta _{ij}$, and setting $%
g_{s}=0$, we can solve (\ref{V}) by 
\begin{eqnarray}
m_{1} &=&0,\quad \\
m_{2} &=&\frac{g_{l}}{\sqrt{3}}\left[ V(\tilde{\alpha}_{2,0}\cdot q)+V(%
\tilde{\alpha}_{2,4}\cdot q)-2V(\tilde{\alpha}_{2,5}\cdot q)\right] , \\
m_{3} &=&\frac{g_{l}}{\sqrt{3}}\left[ 2V(\tilde{\alpha}_{2,4}\cdot q)-V(%
\tilde{\alpha}_{2,0}\cdot q)-V(\tilde{\alpha}_{2,5}\cdot q)\right] .
\end{eqnarray}
To find a general Lax operator which involves all terms of the potential one
needs to device other techniques.

\subsection{The $G_{2}$-Lax operator from double reduction}

The construction procedure is summarized by the following Dynkin diagrams:

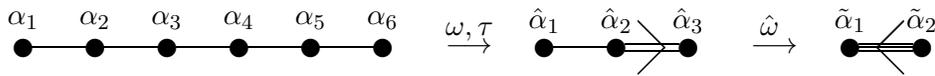
\unitlength=0.680000pt 
\begin{picture}(300.0,70.00)(250.00,125.00)

\put(455.00,165.00){\makebox(0.00,0.00){${{\alpha} }_{6}$}}
\put(415.00,165.00){\makebox(0.00,0.00){${{\alpha} }_{5}$}}
\put(375.00,165.00){\makebox(0.00,0.00){${{\alpha} }_{4}$}}
\put(335.00,165.00){\makebox(0.00,0.00){${{\alpha} }_{3}$}}
\put(295.00,165.00){\makebox(0.00,0.00){${{\alpha} }_{2}$}}
\put(255.00,165.00){\makebox(0.00,0.00){${{\alpha} }_{1}$}}

\put(420.00,150.00){\line(1,0){30.00}}
\put(380.00,150.00){\line(1,0){30.00}}
\put(340.00,150.00){\line(1,0){30.00}}
\put(300.00,150.00){\line(1,0){30.00}}
\put(260.00,150.00){\line(1,0){30.00}}

\put(455.00,150.00){\circle*{10.00}}
\put(415.00,150.00){\circle*{10.00}}
\put(375.00,150.00){\circle*{10.00}}
\put(335.00,150.00){\circle*{10.00}}
\put(295.00,150.00){\circle*{10.00}}
\put(255.00,150.00){\circle*{10.00}}

\put(490.00,157.00){${\omega}, \tau$}
\put(490.00,147.00){$ \longrightarrow $}
\put(545.00,150.00){\circle*{10.00}}
\put(585.00,150.00){\circle*{10.00}}
\put(625.00,150.00){\circle*{10.00}}

\put(550.00,150.00){\line(1,0){30.00}}
\put(590.00,148.00){\line(1,0){30.00}}
\put(590.00,152.00){\line(1,0){30.00}}

\put(612.00,150.00){\line(-1,-1){15.00}}
\put(612.00,150.00){\line(-1,1){15.00}}

\put(545.00,165.00){\makebox(0.00,0.00){${\hat{\alpha} }_{1}$}}
\put(585.00,165.00){\makebox(0.00,0.00){${\hat{\alpha} }_{2}$}}
\put(625.00,165.00){\makebox(0.00,0.00){${\hat{\alpha} }_{3}$}}

\put(665.00,157.00){$ {\hat{\omega}}$}
\put(660.00,147.00){$ \longrightarrow $}
\put(715.00,150.00){\circle*{10.00}}
\put(755.00,150.00){\circle*{10.00}}

\put(720.00,152.00){\line(1,0){30.00}}
\put(720.00,150.00){\line(1,0){30.00}}
\put(720.00,148.00){\line(1,0){30.00}}

\put(730.00,150.00){\line(1,1){15.00}}
\put(730.00,150.00){\line(1,-1){15.00}}

\put(715.00,165.00){\makebox(0.00,0.00){${\tilde{\alpha} }_{1}$}}
\put(755.00,165.00){\makebox(0.00,0.00){${\tilde{\alpha} }_{2}$}}

\end{picture}

\noindent In the first step we start by folding the $A_{6}^{(1)}$-root
system to the $A_{6}^{(2)}$-root system by means of a map $\omega $.
Subsequently we constrain some of the coupling constants through a map $\tau 
$, which amounts to an elimination of some particular roots of $A_{6}$, such
that we obtain two copies of a $B_{3}$-root system. From this system we
obtain the $G_{2}$-root system, by the action of a further map $\hat{\omega}$%
.

The L-operator for the $G_{2}$-Calogero-Moser model then reads 
\begin{eqnarray}
L(\tilde{p},\tilde{q}) &=&\omega ^{-1}\hat{\omega}^{-1}\tilde{p}\cdot
H+\sum\limits_{i=1}^{6}\sum\limits_{p=0}^{6}\tau (g_{i,p})f(\hat{\omega}%
\omega \alpha _{i,p}\cdot \tilde{q})E_{\alpha _{i,p}}  \label{L} \\
&=&\omega ^{-1}\hat{\omega}^{-1}\tilde{p}\cdot
H+\sum\limits_{i=1}^{6}\sum\limits_{p=0}^{6}\tau (g_{i,p})f(\alpha
_{i,p}\cdot \omega ^{-1}\hat{\omega}^{-1}\tilde{q})E_{\alpha _{i,p}}
\end{eqnarray}
with $H_{i},E_{\alpha _{i,p}}\in A_{6}$. We shall now specify the maps $%
\omega $, $\hat{\omega}$, $\tau $ in detail, construct the corresponding $M$%
-operator and show that the Lax equation holds upon the validity of the $%
G_{2}$-equation of motion.

\subsubsection{Reduction of the root systems}

Let us precisely see how the root systems are embedded into each other as $%
\tilde{\Delta}_{G_{2}}\supset $ $\hat{\Delta}_{B_{3}}\supset \Delta _{A_{6}}$%
. We label the 42 roots of$\ A_{6}$-are as

\begin{center}
$ 
\begin{tabular}{|c|ccccc|}
\hline
$i\diagdown p\!$ & \multicolumn{1}{||c}{0} & \multicolumn{1}{|c}{1} & 
\multicolumn{1}{|c}{2} & \multicolumn{1}{|c}{3} & \multicolumn{1}{|c|}{4} \\ 
\hline\hline
1 & \multicolumn{1}{|r|}{$\alpha _{1}\!$} & \multicolumn{1}{r|}{\underline{$%
\alpha _{2}+\alpha _{3}$}$\!$} & \multicolumn{1}{r|}{\underline{$\alpha
_{4}+\alpha _{5}$}$\!$} & \multicolumn{1}{r|}{$\alpha _{6}\!$} & 
\multicolumn{1}{r|}{} \\ \hline
2 & \multicolumn{1}{|r|}{$\!\!-\alpha _{2}\!$} & \multicolumn{1}{r|}{%
\underline{$\alpha _{1}+\alpha _{2}+\alpha _{3}$}$\!$} & \multicolumn{1}{r|}{%
$\mathbf{\alpha }_{2}\mathbf{+\alpha }_{3}\mathbf{+\alpha }_{4}\mathbf{%
+\alpha }_{5}\!$} & \multicolumn{1}{r|}{\underline{$\alpha _{4}+\alpha
_{5}+\alpha _{6}$}$\!$} & \multicolumn{1}{r|}{$-\alpha _{5}\!$} \\ \hline
3 & \multicolumn{1}{|r|}{\underline{$\alpha _{3}$}$\!$} & 
\multicolumn{1}{r|}{$\alpha _{1}+\alpha _{2}+\alpha _{3}+\alpha _{4}+\alpha
_{5}\!$} & \multicolumn{1}{r|}{$\alpha _{2}+\alpha _{3}+\alpha _{4}+\alpha
_{5}+\alpha _{6}\!$} & \multicolumn{1}{r|}{$\alpha _{4}\!$} & 
\multicolumn{1}{r|}{} \\ \hline
4 & \multicolumn{1}{|r|}{\underline{$-\alpha _{4}$}$\!$} & 
\multicolumn{1}{r|}{$\alpha _{3}+\alpha _{4}+\alpha _{5}\!$} & 
\multicolumn{1}{r|}{$\mathbf{\alpha }_{1}\mathbf{+\alpha }_{2}\mathbf{%
+\alpha }_{3}\mathbf{+\alpha }_{4}\mathbf{+\alpha }_{5}\mathbf{+\alpha }%
_{6}\!$} & \multicolumn{1}{r|}{$\alpha _{2}+\alpha _{3}+\alpha _{4}\!$} & 
\multicolumn{1}{r|}{\underline{$-\alpha _{3}$}$\!$} \\ \hline
5 & \multicolumn{1}{|r|}{$\alpha _{5}\!$} & \multicolumn{1}{r|}{$\alpha
_{3}+\alpha _{4}+\alpha _{5}+\alpha _{6}\!$} & \multicolumn{1}{r|}{$\alpha
_{1}+\alpha _{2}+\alpha _{3}+\alpha _{4}\!$} & \multicolumn{1}{r|}{$\alpha
_{2}\!$} & \multicolumn{1}{r|}{} \\ \hline
6 & \multicolumn{1}{|r|}{$\!\!-\alpha _{6}\!$} & \multicolumn{1}{r|}{$\alpha
_{5}+\alpha _{6}\!$} & \multicolumn{1}{r|}{$\mathbf{\alpha }_{3}\mathbf{%
+\alpha }_{4}\!$} & \multicolumn{1}{r|}{$\alpha _{1}+\alpha _{2}\!$} & 
\multicolumn{1}{r|}{$-\alpha _{1}\!$} \\ \hline
\end{tabular}
\ \medskip $

The roots $\hat{\alpha}_{i,p}$ of the $\hat{\Delta}_{A_{6}}$-root system.
\end{center}

We did not report values of $p\geq 4$, i.e. powers of the Coxeter element,
for those roots which can be obtained simply by a multiplication with $-1$
from a root of another orbit of the Coxeter element. For instance $\alpha
_{1,4}=-\alpha _{6,1}$, $\alpha _{1,5}=-\alpha _{6,2}$, etc. Let us now
specify the action of the folding map $\omega $, which acts on the simple
roots of $A_{6}$ 
\begin{equation}
\alpha _{i}\mapsto \omega (\alpha _{i})=\left\{ 
\begin{array}{ll}
\hat{\alpha}_{i\!} & \text{for \ }i=1,2,3 \\ 
\hat{\alpha}_{7-i}\qquad & \text{for \ }i=4,5,6.%
\end{array}
\right.  \label{oh}
\end{equation}
Comparing the entire root system resulting in this manner with the $(\ell
=3)\times (h=6)=18$ roots of $B_{3}^{(1)}$

\begin{center}
$ 
\begin{tabular}{|c||cccccc|}
\hline
$i\diagdown p$ & 0 & \multicolumn{1}{|c}{1} & \multicolumn{1}{|c}{2} & 
\multicolumn{1}{|c}{3} & \multicolumn{1}{|c|}{4} & 5 \\ \hline\hline
1 & \multicolumn{1}{|r|}{$\hat{\alpha}_{1\!}$} & \multicolumn{1}{r|}{$\hat{%
\alpha}_{2}+2\hat{\alpha}_{3}\!$} & \multicolumn{1}{r|}{$\hat{\alpha}_{1}+%
\hat{\alpha}_{2}\!$} & \multicolumn{1}{r|}{$-\hat{\alpha}_{1}\!$} & 
\multicolumn{1}{r|}{$-(\hat{\alpha}_{2}+2\hat{\alpha}_{3})\!$} & $-(\hat{%
\alpha}_{1}+\hat{\alpha}_{2})$ \\ \hline
2 & \multicolumn{1}{|r|}{$-\hat{\alpha}_{2}\!$} & \multicolumn{1}{r|}{$\!\!%
\hat{\alpha}_{1}+\hat{\alpha}_{2}+2\hat{\alpha}_{3}\!$} & 
\multicolumn{1}{r|}{$\!\!\!\hat{\alpha}_{1}+2\hat{\alpha}_{2}+2\hat{\alpha}%
_{3\!}$} & \multicolumn{1}{r|}{$\hat{\alpha}_{2}\!$} & \multicolumn{1}{r|}{$%
\!\!\!-(\hat{\alpha}_{1}+\hat{\alpha}_{2}+2\hat{\alpha}_{3})\!$} & $\!\!\!-(%
\hat{\alpha}_{1}+2\hat{\alpha}_{2}+2\hat{\alpha}_{3})\!\!$ \\ \hline
3 & \multicolumn{1}{|r|}{$\hat{\alpha}_{3}\!$} & \multicolumn{1}{r|}{$\hat{%
\alpha}_{1}+\hat{\alpha}_{2}+\hat{\alpha}_{3}\!$} & \multicolumn{1}{r|}{$%
\hat{\alpha}_{2}+\hat{\alpha}_{3}\!$} & \multicolumn{1}{r|}{$-\hat{\alpha}%
_{3}\!$} & \multicolumn{1}{r|}{$-(\hat{\alpha}_{1}+\hat{\alpha}_{2}+\hat{%
\alpha}_{3})\!$} & $-(\hat{\alpha}_{2}+\hat{\alpha}_{3})$ \\ \hline
\end{tabular}
\ \medskip $

The roots $\hat{\alpha}_{i,p}$ of the $\hat{\Delta}_{B_{3}}$-root system.
\end{center}

\noindent it is easy to see that the map (\ref{oh}) reduces the $A_{6}$-root
system to two copies of a $B_{3}$-root system plus 6 additional roots. We
marked the 3 positive roots in the table of $\Delta _{A_{6}}$, which are not
mapped to $\hat{\Delta}_{B_{3}}$ via $\omega $ in bold and underlined the
roots which are mapped to short roots in $\hat{\Delta}_{B_{3}}$. The
unmarked roots are therefore mapped to long roots.

Having specified the map $\omega :$ $\Delta _{A_{6}}\rightarrow $ $\hat{%
\Delta}_{B_{3}}$ acting on the roots, it is important to see how this
reduction is translated to the action on the dynamical variables.\ For this
purpose\ we construct its \textquotedblleft inverse\textquotedblright\ map\ $%
\omega ^{-1}:$ $\hat{\Delta}_{B_{3}}\rightarrow $ $\Delta _{A_{6}}$, defined
via the inner product relation 
\begin{equation}
\omega (\alpha _{i})\cdot \hat{\alpha}_{j}=\alpha _{i}\cdot \omega ^{-1}(%
\hat{\alpha}_{j}),\qquad \text{\quad\ }1\leq i\leq 6,1\leq j\leq 3.
\end{equation}
It is easy to verify that this is guaranteed by the map 
\begin{equation}
\hat{\alpha}_{i}\mapsto \omega ^{-1}(\hat{\alpha}_{i})=\alpha _{i}+\alpha
_{7-i},\qquad \text{\quad\ }1\leq i\leq 3,
\end{equation}
when taking the conventions $\alpha ^{2}=2$, $\hat{\alpha}_{1}^{2}=\hat{%
\alpha}_{2}^{2}=2$ and $\hat{\alpha}_{3}^{2}=1$. Now we may utilize this map
to compute the reduction map when acting on the dynamical variables $q,p$%
\begin{equation}
q\rightarrow \omega ^{-1}(\hat{q})=\omega ^{-1}\left(
\dsum\nolimits_{i=1}^{3}\hat{y}_{i}\hat{\alpha}_{i}\right)
=(y_{1},y_{2}-y_{1},y_{3}-y_{2},0,y_{2}-y_{3},y_{1}-y_{2},-y_{1}),  \label{q}
\end{equation}
where we used the aforementioned Euclidean realization for the $A_{6}$ root
system in $\mathbb{R}^{7}$. To make contact with the literature, we defined
a further set of variables through the relation $y_{i}=$ $%
\dsum\nolimits_{k=1}^{i}\hat{q}_{k}$, such that (\ref{q}) becomes 
\begin{equation}
q\rightarrow (\hat{q}_{1},\hat{q}_{2},\hat{q}_{3},0,-\hat{q}_{3},-\hat{q}%
_{2},-\hat{q}_{1}).
\end{equation}
This is the reduction map as employed in \cite{OP6} (see also \cite{Per}).

Likewise, we reduce next the $B_{3}$-root system to the $G_{2}$-root system
by means of the map $\hat{\omega}:$ $\hat{\Delta}_{B_{3}}\rightarrow $ $%
\tilde{\Delta}_{G_{2}}$ 
\begin{equation}
\hat{\alpha}_{i}\mapsto \hat{\omega}(\hat{\alpha}_{i})=\left\{ 
\begin{array}{ll}
\tilde{\alpha}_{1\!} & \text{for \ }i=1,3 \\ 
\tilde{\alpha}_{2\!}\qquad  & \text{for \ }i=2.%
\end{array}%
\right. 
\end{equation}%
The \textquotedblleft inverse\textquotedblright\ $\hat{\omega}^{-1}:$ $%
\tilde{\Delta}_{G_{2}}\rightarrow $ $\hat{\Delta}_{B_{3}}$ is obtained \
similarly as before, but now demanding 
\begin{equation}
\hat{\omega}(\hat{\alpha}_{i})\cdot \tilde{\alpha}_{j}=\hat{\alpha}_{i}\cdot 
\hat{\omega}^{-1}(\tilde{\alpha}_{j}),\qquad \text{\quad\ }1\leq i\leq
3,1\leq j\leq 2.
\end{equation}%
We find 
\begin{equation}
\tilde{\alpha}_{1}\mapsto \hat{\omega}^{-1}(\tilde{\alpha}_{1})=\hat{\alpha}%
_{1}+2\hat{\alpha}_{3}\quad \text{and\quad }\tilde{\alpha}_{2}\mapsto \hat{%
\omega}^{-1}(\tilde{\alpha}_{2})=3\hat{\alpha}_{2},
\end{equation}%
with the additional conventions $\tilde{\alpha}_{1}^{2}=2$ and $\tilde{\alpha%
}_{2}^{2}=6$. The reduction map, when acting on the dynamical variables $%
\hat{q},\hat{p}$, is now evaluated as 
\begin{equation}
\hat{q}\rightarrow \omega ^{-1}(\tilde{q})=\omega ^{-1}\left(
\dsum\nolimits_{i=1}^{2}\tilde{y}_{i}\hat{\alpha}_{i}\right) =(\tilde{y}%
_{1},3\tilde{y}_{2}-\tilde{y}_{1},2\tilde{y}_{1}-3\tilde{y}_{2})=(-\tilde{q}%
_{1}^{^{\prime }},\tilde{q}_{2}^{^{\prime }},\tilde{q}_{3}^{^{\prime }}),
\end{equation}%
where we realized the $B_{3}$-roots in $\mathbb{R}^{3}$ as $\hat{\alpha}%
_{1\!}=\varepsilon _{1}-\varepsilon _{2}$, $\hat{\alpha}_{2}=\varepsilon
_{2}-\varepsilon _{3}$ and $\hat{\alpha}_{3}=\varepsilon _{3}$. The
introduction of the variables $\tilde{q}_{i}^{^{\prime }}$ translates into
the usual $G_{2}$ constraint $\tilde{q}_{1}+\tilde{q}_{2}+\tilde{q}_{3}=0$,
which corresponds to considering the three particle system in the center of
mass frame. In fact, it will be convenient to introduce yet another set of
variables, namely $\tilde{q}_{1}^{^{\prime }}=\tilde{q}_{2}-\tilde{q}_{3}$, $%
\tilde{q}_{2}^{^{\prime }}=\tilde{q}_{3}-\tilde{q}_{1}$ and $\tilde{q}%
_{3}^{^{\prime }}=\tilde{q}_{1}-\tilde{q}_{2}$ to make proper contact with (%
\ref{v2}).

Let us now see how to utilize these maps in order to reduce the
corresponding potentials.

\subsubsection{Reduction of the potentials}

Our starting point is the $A_{6}$-potential term in the form 
\begin{equation}
V_{A_{6}}(q)=\frac{1}{2}\dsum\limits_{i=1}^{6}\dsum\limits_{p=0}^{6}g_{i%
\!,p}^{2}V(\alpha _{i\!,p}\cdot q).~
\end{equation}%
Below we confirm the known fact that the Lax equation dictates that all
coupling constants have to be taken to be the same, i.e. $g_{i\!,p}=g$. Then 
$V_{A_{6}}$ is mapped into a $B_{3}$-potential of the form 
\begin{equation}
V_{B_{3}}(\hat{q})=\frac{1}{2}\dsum\limits_{i=1}^{6}\dsum\limits_{p=0}^{6}%
\tau (g_{i\!,p})^{2}V(\omega \alpha _{i\!,p}\cdot \hat{q})=\frac{\hat{g}^{2}%
}{2}\dsum\limits_{p=0}^{5}\left[ \dsum\limits_{i=1}^{2}V(\hat{\alpha}%
_{i,p}\cdot \hat{q})+2V(\hat{\alpha}_{3,p}\cdot \hat{q})\right] ,  \label{B3}
\end{equation}%
by means of the reduction map $\omega $ as specified above (\ref{oh}) and
the map $\tau $ acting on the coupling constants as 
\begin{equation}
g_{i\!,p}\mapsto \tau (g_{i\!,p})=\left\{ \!\!\!%
\begin{array}{ll}
0 & \text{for }g_{1,5},g_{3,5},g_{5,5},g_{2,2},g_{4,2},g_{6,2} \\ 
\hat{g} & \text{for }%
g_{1,1},g_{1,2},g_{2,1},g_{2,3},g_{3,0},g_{3,3},g_{4,0},g_{4,4},g_{5,4},g_{5,6},g_{6,5},g_{6,6}
\\ 
\frac{\hat{g}}{\sqrt{2}}~ & \text{otherwise}.%
\end{array}%
\right.   \label{gg}
\end{equation}%
The map $\tau $ serves here to eliminate the aforementioned additional six
roots of $A_{6}$ which have no counterpart in $B_{3}$ and at the same time
it establishes a relationship\ between the coupling constants depending on
whether the potential involves roots which are mapped to long or short
roots. This relation is dictated by the Lax pair construction, and coincides
with the one found by Olshanetsky and Perelomov in \cite{OP6}, or \cite{Per}
p. 181. Note that taking merely the invariance of the Coxeter transformation
as a guiding principle one could choose the coupling constants in front of
the term involving long or short roots to be independent, see e.g.
discussion in \cite{FK}. However, integrability demands the dependence of
the coupling constants to be as stated in\ (\ref{gg}), such that one has
only one coupling constant at ones disposal for the $B_{\ell }$-theories.

Next we map the $B_{3}$-potential to the $G_{2}$-potential with the help of $%
\hat{\omega}$ and find 
\begin{equation}
\!V_{G_{2}}\!=\!\frac{\tilde{g}^{2}}{2}\dsum\limits_{p=0}^{5}\left[
\dsum\limits_{i=1}^{2}V(\hat{\omega}\hat{\alpha}_{i,p}\cdot \tilde{q})+2V(%
\hat{\omega}\hat{\alpha}_{3,p}\cdot \tilde{q})\right] =\frac{\tilde{g}^{2}}{2%
}\dsum\limits_{p=0}^{5}\left[ 3V(\tilde{\alpha}_{1\!,p}\cdot \tilde{q})+V(%
\tilde{\alpha}_{3\!,p}\cdot \tilde{q})\right] .
\end{equation}
In this last reduction step we did not need to specify any additional map
acting on the coupling constants as the embedding is now on-to-one. For
consistency we re-named, however, $\hat{g}$ to $\tilde{g}$.

\subsubsection{Constraints from the Lax operator}

Having convinced ourselves that the potentials can be reduce properly, we
still have to establish that the Lax operator exists and is indeed of the
form (\ref{L}). We commence by explicitly solving the constraint (\ref{V})
for $A_{6}$. In principle there are now $42\times 42$ possible structure
constants $\varepsilon _{i,p,j,q}$, with 420 of them non-vanishing. We only
report here our conventions for the signs of the 35 essentials and leave it
to the reader to obtain the remaining ones by means of the equations (\ref%
{eps1}) and (\ref{eps2}) 
\begin{eqnarray}
\varepsilon _{1,0,1,1} &=&\varepsilon _{1,0,2,2}=\varepsilon
_{1,0,3,2}=\varepsilon _{1,1,1,2}=\varepsilon _{1,1,2,3}=\varepsilon
_{1,1,3,3}=\varepsilon _{1,2,1,3}=\varepsilon _{1,2,2,4}=\varepsilon
_{1,3,1,4}=\quad \quad ~~~  \label{e1} \\
\varepsilon _{1,3,2,5} &=&\varepsilon _{1,3,3,5}=\varepsilon
_{1,4,1,5}=\varepsilon _{1,4,2,6}=\varepsilon _{1,5,1,6}=\varepsilon
_{1,5,2,0}=\varepsilon _{1,5,3,0}=\varepsilon _{1,6,1,0}=\varepsilon
_{1,6,2,1}= \\
\varepsilon _{1,6,3,1} &=&\varepsilon _{2,0,1,1}=\varepsilon
_{2,0,2,2}=\varepsilon _{2,1,1,2}=\varepsilon _{2,1,2,3}=\varepsilon
_{2,2,1,3}=\varepsilon _{2,2,2,4}=\varepsilon _{2,3,1,4}=\varepsilon
_{2,3,2,5}= \\
\varepsilon _{2,4,1,5} &=&\varepsilon _{2,4,2,6}=\varepsilon
_{2,5,1,6}=\varepsilon _{2,5,2,0}=\varepsilon _{2,6,1,0}=\varepsilon
_{2,6,2,1}=\varepsilon _{3,0,1,2}=\varepsilon _{3,2,1,4}=1.  \label{e4}
\end{eqnarray}%
In fact, we verified that these choices coincide with the constants obtained
directly from (\ref{comm}) when using the vector representation of $A_{6}$.
With (\ref{e1})-(\ref{e4}) and the above mentioned realization for the
simple roots, the constraint\ (\ref{V}) may be solved by 
\begin{equation}
m_{i}^{A_{6}}=g\sum_{\substack{ k=1 \\ k\neq i}}^{7}V_{A_{6}}(q_{k}-q_{i}),
\end{equation}%
which is known for some time \cite{Per}. Next we may solve (\ref{V}) for the
reduced systems and find 
\begin{equation}
m_{i}^{B_{3}}=\sqrt{2}\hat{g}\sum_{\substack{ k=1 \\ k\neq i}}^{7}\tau
_{ki}V_{A_{6}}(\hat{\omega}^{-1}(\hat{q}_{k})-\hat{\omega}^{-1}(\hat{q}_{i}))
\end{equation}%
with $\tau _{ki}=1$ except for $\tau _{4i}=2$, $\tau _{(8-i)i}=0$ and 
\begin{equation}
m_{i}^{G_{2}}=\sqrt{2}g\sum_{\substack{ k=1 \\ k\neq i}}^{7}\tau
_{ki}V_{A_{6}}(\omega ^{-1}\hat{\omega}^{-1}(\tilde{q}_{k})-\omega ^{-1}\hat{%
\omega}^{-1}(\tilde{q}_{i})).
\end{equation}%
Having presented explicit solutions to the equation (\ref{V}), we have
established the existence of the operators $L$ and $M$. In particular (\ref%
{L}) and the corresponding equation for $M$ satisfy the Lax equation up to
the validity of the $G_{2}$ equations of motion.

\subsubsection{Conserved Charges}

It is instructive to consider an explicit matrix representation for the
L-operator.  Using the standard vector representation of $A_{6}$ it follows
directly form (\ref{L}) 
\begin{equation}
\!L=\frac{\lambda }{\sqrt{2}}\left( 
\begin{array}{ccccccc}
\frac{\sqrt{2}}{\lambda }\tilde{p}_{32} & f(\tilde{q}_{12}) & f(\tilde{q}%
_{31}) & \sqrt{2}f(\tilde{q}_{32}) & f(\tilde{q}_{13,2}) & f(\tilde{q}%
_{3,12}) & 0 \\ 
f(\tilde{q}_{21}) & \frac{\sqrt{2}}{\lambda }\tilde{p}_{31} & f(\tilde{q}%
_{23,1}) & \sqrt{2}f(\tilde{q}_{31}) & f(\tilde{q}_{32}) & 0 & f(\tilde{q}%
_{3,12}) \\ 
f(\tilde{q}_{13}) & f(\tilde{q}_{1,23}) & \frac{\sqrt{2}}{\lambda }\tilde{p}%
_{12} & \sqrt{2}f(\tilde{q}_{12}) & 0 & f(\tilde{q}_{32}) & f(\tilde{q}%
_{13,2}) \\ 
\sqrt{2}f(\tilde{q}_{23}) & \sqrt{2}f(\tilde{q}_{13}) & \sqrt{2}f(\tilde{q}%
_{21}) & 0 & \sqrt{2}f(\tilde{q}_{12}) & \sqrt{2}f(\tilde{q}_{31}) & \sqrt{2}%
f(\tilde{q}_{32}) \\ 
f(\tilde{q}_{2,13}) & f(\tilde{q}_{23}) & 0 & \sqrt{2}f(\tilde{q}_{21}) & 
\frac{\sqrt{2}}{\lambda }\tilde{p}_{21} & f(\tilde{q}_{23,1}) & f(\tilde{q}%
_{31}) \\ 
f(\tilde{q}_{12,3}) & 0 & f(\tilde{q}_{23}) & \sqrt{2}f(\tilde{q}_{13}) & f(%
\tilde{q}_{1,23}) & \frac{\sqrt{2}}{\lambda }\tilde{p}_{13} & f(\tilde{q}%
_{12}) \\ 
0 & f(\tilde{q}_{12,3}) & f(\tilde{q}_{2,13}) & \sqrt{2}f(\tilde{q}_{23}) & 
f(\tilde{q}_{13}) & f(\tilde{q}_{21}) & \frac{\sqrt{2}}{\lambda }\tilde{p}%
_{23}%
\end{array}%
\right) 
\end{equation}%
where we abbreviated $\tilde{p}_{ij}:=\tilde{p}_{i}-\tilde{p}_{j}$, $\tilde{q%
}_{ij}:=\tilde{q}_{i}-\tilde{q}_{j}$, $\tilde{q}_{ij,k}:=\tilde{q}_{i}+%
\tilde{q}_{j}-2\tilde{q}_{k}$ and $\tilde{q}_{k,ij}:=2\tilde{q}_{k}-\tilde{q}%
_{i}-\tilde{q}_{j}$. By simple matrix multiplication we compute from this
the integrals of motion of the form $I_{k}=\func{tr}(L^{k})/k$ 
\begin{eqnarray}
I_{1} &=&0,\quad I_{2}=\mathcal{H},\quad I_{3}=0,\quad I_{4}=\frac{1}{4}%
I_{2}^{2},\quad I_{5}=0,\quad I_{6}\neq 0,\quad I_{7}=0, \\
I_{8} &=&I_{2}I_{6}-\frac{5}{96}I_{2}^{4},\quad I_{9}=0,\quad I_{10}=\frac{3%
}{4}I_{2}^{2}I_{6}-\frac{1}{20}I_{2}^{5},\quad I_{11}=0,\quad  \\
I_{12} &=&\frac{5}{12}I_{2}^{3}I_{6}+\frac{1}{2}I_{6}^{2}-\frac{19}{576}%
I_{2}^{6},\quad I_{13}=0,\ldots 
\end{eqnarray}%
Thus we find non-vanishing and algebraically independent charges $I_{k}$
only for $k$ being a degree of $G_{2}$, that is 2 and 6, see e.g. \cite{Hum2}%
. Computing the $L$-operator of the $B_{3}$-Calogero-Moser model we verify
the same property.

\section{Conclusions}

We have constructed a simple expression for the $L$-operator of the $G_{2}$%
-Calogero-Moser model. We established that the constraint (\ref{V}) may
indeed be solved and therefore that the $L$ and $M$-operator do exist. The
operators are expanded in terms of $H,E_{\alpha }\in A_{6}$. The Lax
equation constructed from these operators, with coefficients subject to the
stated reduction maps, holds up to the validity of the $G_{2}$ equations of
motion. To find a solution to (\ref{V}) and thus guaranteeing the
integrability of the model, we found that we are only permitted to have one
coupling constant in the $G_{2}$-theory, instead of two, what might be
expected from demanding invariance under the Coxeter group. In our approach
this feature is inherited from the $B_{3}$-theory. Such a behaviour was also
observed in \cite{Hok}. It would be interesting to investigate if this
limitation can be overcome by other techniques or to establish that this is
really an intrinsic feature of the model. Furthermore, from the explicit
computations of numerous integrals of motion, we found that they are only
algebraically independent and non-vanishing if their degree is equal to the
degree of the corresponding Lie algebra.

In our discussion we did not appeal to the explicit form of the potential
and only require the relation (\ref{f}) to be satisfied. This means that the
models covered here are of the general form $V(x)\sim 1/\func{sn}^{2}(x)$
including a dependence of spectral parameter $\mu $. It would be interesting
to investigate the properties of the spectral curves $R(k,\mu )=\det \left[ k%
\mathbb{I}-L(\mu )\right] $ in the spirit of \cite{Krich,BB,Don,Hok},
resulting from the $L$-operator presented here.\medskip 

\noindent \textbf{Acknowledgments}. A.F. is grateful to the Departamento de
Matem\'{a}tica, F.C.T. Universidade do Algarve, for extremely kind
hospitality and to Christian Korff for many useful discussions. This work
was supported by the FCT project POCI/MAT/58452/2004.


\end{document}